# Visualizing Individual Nitrogen Dopants in Monolayer Graphene

Liuyan Zhao[1], Rui He[1], Kwang Taeg Rim[2], Theanne Schiros[3], Keun Soo Kim[1,4], Hui Zhou[1], Christopher Gutiérrez[1], S. P. Chockalingam[1], Carlos J. Arguello[1],
Lucia Pálová[2], Dennis Nordlund[5], Mark S. Hybertsen[6],
David R. Reichman[2], Tony F. Heinz[1,7], Philip Kim[1], Aron Pinczuk[1,8], George W. Flynn[2], Abhay N. Pasupathy[1]

[1]Department of Physics, Columbia University, New York NY 10027, USA
[2]Department of Chemistry, Columbia University, New York NY 10027, USA
[3]Energy Frontier Research Center, Columbia University, New York NY 10027, USA
[4]Department of Physics and Graphene Research Institute, Sejong University, Seoul 143-747, Korea
[5]Stanford Synchrotron Radiation Lightsource, SLAC National Accelerator Laboratory, Menlo Park, CA 94025, USA
[6]Center for Functional Nanomaterials, Brookhaven National Laboratory, Upton, New York, 11973-5000, USA
[7]Department of Electrical Engineering, Columbia University, New York NY 10027, USA
[8]Department of Applied Physics and Applied Mathematics, Columbia University, New York NY 10027, USA

In monolayer graphene, substitutional doping during growth can be used to alter its electronic properties. We used scanning tunneling microscopy (STM), Raman spectroscopy, x-ray spectroscopy, and first principles calculations to characterize individual nitrogen dopants in monolayer graphene grown on a copper substrate. Individual nitrogen atoms were incorporated as graphitic dopants, and a fraction of the extra electron on each nitrogen atom was delocalized into the graphene lattice. The electronic structure of nitrogen-doped graphene was strongly modified only within a few lattice spacings of the site of the nitrogen dopant. These findings show that chemical doping is a promising route to achieving high-quality graphene films with a large carrier concentration.

Substitutional doping is a powerful way to tailor material properties at the nanoscale (*1*), and it might be expected to have fundamentally different consequences when used to alter the electronic properties of inherently two-dimensional (2D) materials such as graphene (*2*). In recent years, several experimental techniques have been developed to dope the carbon lattice (*3-9*). These include methods applied during growth of large-area graphene films (*4-5, 7-8*) and ways to modify the material after growth (*3,6*), as well as a new one-pot procedure to produce highly doped, few layer graphitic structures (*9*). Several characterization techniques including x-ray photoemission (*10-11*), Raman spectroscopy (*3, 5, 7*) and transmission-electron microscopy (*12*) have been used to analyze the effect of the doping process in graphene. However, a microscopic understanding of the atomic and low-energy electronic structure induced by the substitutional doping process in monolayer graphene is lacking. In our experiment, we use the atomic-resolution imaging capabilities of the scanning tunneling microscope to understand the local structure in the vicinity of a N dopant in monolayer graphene, and spectroscopic imaging to measure the density of states and carrier concentration at the nanoscale. We also report scanning

Raman spectroscopy, core-level x-ray spectroscopy and first principles calculations to characterize the effect of N-doping on the graphene film.

We grew N-doped graphene films using chemical vapor deposition on copper foil substrates (*13*) in a quartz tube furnace [sample growth details in supporting online material (SOM) I(a)]. Each graphene-covered foil was divided into three parts and immediately studied using Raman spectroscopy, x-ray spectroscopy and scanning tunneling microscopy (STM) measurements. Samples have been characterized as grown on copper foils as well as after transfer to an insulating $SiO_2$ substrate [sample transfer and preparation details in SOM I, (b) to (d)].

Raman spectra (Fig. 1A) were taken over large areas (20 μm spot size) of the N-doped samples on $SiO_2$ substrates and on a pristine graphene sample for reference. Raman spectra taken on copper foil substrates showed identical features along with a luminescence background from the copper [see figure S1 and SOM II(a)]. The pristine graphene spectrum showed the G and 2D resonances at 1585.5 and 2684.5 $cm^{-1}$ (*14*); little or no signal at the D band at 1348.5 $cm^{-1}$ was observed, indicating a low defect density. In contrast, the doped samples exhibit strong D and D' resonances (*6, 15-17*), as well as changes in the intensity and position of the G and 2D bands (*18-19*). The Raman spectra on N-doped graphene were consistent with the presence of dopants or defects that modify the free carrier concentration while preserving the basic structural properties of the graphene sheet.

We confirmed the presence of N dopants in the graphene lattice by performing x-ray photoelectron spectroscopy (XPS; N1s) and near edge x-ray absorption fine structure (NEXAFS) spectroscopy at the nitrogen K-edge for a pristine and NG10 doped sample (Fig. 1B) [see SOM I(e)]. The addition of $NH_3$ in the graphene growth resulted in sharp peaks at 400.7 and 408 eV in the NEXAFS spectrum, corresponding to 1s to π* and σ* transitions, respectively, for a single molecular species. The sharpness and strong polarization dependence of the peaks indicates that this species has well-defined, in-plane N-C bonds in the graphene lattice. Based on previous studies (*20-22*), this peak can be assigned to $sp^2$-bonded graphitic nitrogen with three C neighbors. Representative N1s XPS shows a higher binding energy component (black line) for the $NH_3$-grown samples, indicating the formation of more electronegative N-C bonds in the graphene lattice as opposed to at edges or defects (gray line). This higher binding energy peak is generally considered a signature of graphitic N in studies of modified carbon films (*6, 21, 23-24*).

For STM measurements, the graphene-coated substrates were transferred in ambient soon after preparation into an ultra-high vacuum (UHV) low-temperature instrument that is capable of picometer resolution. The treated copper substrates were degassed in UHV at a temperature of 350 °C for several hours before STM experiments were performed. Although many areas of the copper foil displayed a rough topography because of polycrystallinity, we occasionally found large areas where atomically flat terraces can be observed. The NG10 samples were optimal for STM studies because there is a reasonable probability of observing such flat and clean areas along with the dopant features described below; our detailed STM measurements described below are performed on several such samples between 50 and 77 K.

A representative STM topography of N-doped graphene on a large copper terrace (Fig. 1C) mostly revealed the pristine graphene lattice, but the image also showed several bright objects (nearly identical in appearance) with lateral dimensions of a few atomic spacings, consistent with previous measurements on doped films (*9, 25*).These bright objects were not seen in pristine graphene films (see below), leading us to associate them with N dopants. By simply counting up the number of bright features and associating each with a single N dopant, we arrived at a N doping concentration per carbon atom of 0.34%. Measurements across several similar samples yielded N concentrations between 0.23 and 0.35%.

A close-up topography of one of these doping features (Fig. 2A) revealed three bright spots forming a triangle. The distance between the bright spots was equal to the graphene lattice constant (2.5 Å). Far away from the doping feature, the honeycomb lattice of graphene was recovered. Following the honeycomb lattice through the doping structure situated each of the three bright spots on a C atom of the same sublattice. An STM line scan through the dopant (Fig. 2A, inset) yielded a maximum apparent out-of-plane height of 0.6±0.2Å and would be consistent with the N atom substituting for a C atom in the plane of the graphene. From sampling areas on several different N-doped graphene samples, we observe that more than 90% of the dopants are of this form. The observed STM image closely matched our simulated STM image (Fig. 2B) computed from the local density of states for graphitic doping where one N atom replaces a single C atom, as well as other recent calculations (*26*).Our STM image simulations were carried out using the Tersoff-Hamann approach (*27*) based on density functional theory (DFT) calculations in the local density approximation for exchange and correlation, with a plane-wave basis set, and norm-conserving pseudopotentials as implemented in the Quantum-ESPRESSO package(*28*) [see SOM II(e)]. The calculated STM image shows the brightest features (red spots) on the nearest neighbor C atoms to N (shown in blue). Visible features extended to several lattice spacings from the N, and the overall triangular symmetry of the image matched the experiment.

The STM observations show that the majority of the doping occurs via graphitic substitution in these samples [see SOM II(b) and figure S2 for examples of other doping forms seen]. Larger area STM images of the samples such as Fig. 2C located 14 dopants, all in the graphitic form, which is consistent with the NEXAFS results. Apart from the local structure around each dopant, we also see long "tails" around each dopant. These arise from inter valley electron scattering induced by the N dopant, and similar features have been seen before in other experiments (*29-30*). The Fourier transform (FFT) of Fig. 2C (inset) shows evidence for this strong scattering. The FFT shows two sets of points arranged in hexagons. Whereas the outer hexagon of points in the FFT corresponds to the atomic lattice, the inner hexagon in this FFT arises from the strong inter-valley scattering induced by N dopants (*29-30*).

Large-area STM images such as those in Fig. 1C and 2C, can give us information on whether clustering of dopants occurs during growth. To analyze this observation, we used STM images to calculate all the N-N distances for each sample. Figure 2D shows a histogram of these distances for several different samples on a log scale. A true random distribution of dopants would imply a quadratic distribution of dopant-to-dopant distances, which we observed in all of our samples down to length scales of a few lattice constants. Nearby dopants preferred to incorporate on the same sub lattice of graphene. Indeed, all of the dopants shown in Fig. 2C are located on the same sub lattice, and evidence for this phenomenon is seen in other samples as well [see SOM II(c) and figure S3].

We performed detailed STM spectroscopy measurements of the differential conductance dI/dV (Fig. 2E) to understand the effect of the N atoms on the low-energy electronic structure of the graphene film (here, dI/dV is the derivative of the current with respect to the voltage obtained using a lock-in amplifier). The most prominent features seen in all of the curves, which have been seen previously on monolayer graphene(*31*), are two depressions near zero bias and near -300 meV relative to the tip. The zero-bias feature occurs when energies are too low to excite the optical phonon mode in graphene monolayers that can enhance the tunneling current (*31*). The -300 meV feature occurs near the Dirac point where the density of states of graphene is low (*31-33*). We thus associated the dip seen at -270 mV in Fig. 2E with the Dirac point for an electron-doped graphene layer. Measurements of the Dirac point over a 40 by 40 nm area of an NG10 sample show a distribution of values (Fig. 3A). We used the energy of the Dirac point measured at each position of a sample to convert to a charge carrier density at that location using the relation for the ideal graphene band structure, namely, $n = \dfrac{E_D^2}{\pi(\hbar v_F)^2}$ (here n is the charge carrier density, $E_D$ is the Dirac point energy and $\hbar$ is Planck's constant *h* divided by 2π) . Taking a value for the Fermi velocity $v_F = 10^6$ m/s, we arrive at an average charge carrier density of (5.42±0.83)×10$^{12}$ electrons per cm$^2$. We can compare this to the N-doping concentration calculated by simply counting the number of observed dopants in the area where the measurements are performed. The observed N doping in this area corresponds to 0.34% N atoms per C atom, or, equivalently, a N atom density of 1.3×10$^{13}$cm$^{-2}$. Together with the charge carrier density measured by STM spectroscopy, this result implies that each graphitic N dopant contributes (on average) ~ 0.42±0.07 mobile carriers to the graphene lattice. We have performed such detailed spectroscopic measurements across different samples with N atom concentrations varying from 0.23 to 0.35% (Fig. 3B) and found a strong correlation between the number of N dopants and the extracted charge carrier density. An independent measurement of the carrier concentration based on the shift in the position of the G peak in the Raman spectrum induced by doping (*18-19*) was 5 ± 1.5×10$^{12}$ cm$^{-2}$ [see SOM II(d) and figure S4].

Our DFT calculations(*28*) provide insight to the electrostatic balance between the N dopants and free carriers in the graphene sheet. Focusing on a single, graphitic N dopant, the projected N density of states (pDOS) on the π -system(Fig. 3C) revealed a resonance caused by the N $p_z$ orbital centered 0.3 eV above the Fermi energy ($E_F$). The pDOS for the C nearest neighbor exhibited a shoulder caused by its electronic coupling to the N, with a reduced shoulder on the next nearest neighbor. The occupied fraction of these resonances represented the localized charge near the N centers. The balancing charge went to the rest of the pi states. The Dirac point still appeared in the pDOS, shifted to below $E_F$, as also seen in our total DOS and in other, recent calculations (*26,34*). We studied N concentrations (from 0.6 to 5.6%) by varying the super cell size for a single dopant. Using the Dirac point shift in the total DOS as a measure of free electron concentration, we estimate (using the formula above) that 50 to 70% of electrons are delocalized, with oscillations indicative of electronic interference effects in the simulation cells that we used [see SOM II(e) and figure S5]. A value of ~ 60% would be consistent with the experiment values shown in Fig. 3B.

To draw comparisons with other measurements of doping in graphene films, it is important to understand the effect of the copper foil substrate on the carrier concentration via charge transfer (*35-36*) as well as by changing the charge screening in the graphene layer. We performed STM measurements of pristine graphene films on copper foils and estimated that the doping induced by the copper foil substrate into the graphene film is < $10^{12}$ electrons/cm$^2$ [see SOM II(f) and figure S7]. The copper substrate can also modify the charge screening length in the graphene film. We studied doped graphene films transferred to a SiO$_2$ dielectric substrate (*37*), a process that left residue on or below the graphene surface. Typical STM and atomic force microscopy images displayed surface roughness of a few nanometers, but occasionally we found small regions of the sample where the graphene honeycomb lattice could be resolved. Figure 3C shows the average dI/dV spectrum taken over one such clean area (the STM image is shown in the inset in the derivative mode to remove the overall roughness of the substrate and enhance the atomic contrast). The overall features in the spectrum are preserved, but now $E_D$ =-330±20 meV ($n$ =8×10$^{12}$cm$^{-2}$). The transfer processes, as well as the SiO$_2$ substrate itself, introduce an unknown doping concentration into the film; thus it is not possible to directly compare the doping level of the two samples. However, the spectrum of doped graphene on SiO$_2$ is broadly consistent with the spectrum of doped graphene on copper foil, indicating the absence of a strong hybridization between the graphene and underlying copper.

The local electronic structure around a N dopant also affects the electronic nature of the doped graphene film. In Fig. 4A, we show spectra obtained on and far away (~ 2 nm) from a dopant atom. Although the overall features of the spectra were preserved on the N atom, the electron-hole asymmetry in the local density of states was much stronger on the N atom, in accord with DFT calculations (Fig. 3B). The amount of scattering introduced by the dopant can be assessed by STM imaging (*29, 32, 38*). We performed STM spectroscopic mapping in a 2.5 nm area around one graphitic N dopant (STM junction parameters are V=0.8V, I=1 nA). Shown in Fig. 4B are a subset of these maps, acquired at bias voltages (applied to the sample relative to the tip) from -0.78 to 0.78V. The maps did not show much contrast at high positive bias but the local DOS around the N atom was strongly suppressed at energies below the $E_F$. The local DOS recovered its background value within a few lattice constants of the dopant atom. We plotted the radial distribution of the dI/dV intensity (Fig. 4C) from the maps in Fig. 4B as a function of distance from the N atom, normalizing the background value of the dI/dV to unity for each energy. The intensity of the spectral weight changes caused by the N dopant were energy dependent, but the decay lengths were ~ 7Å for all energies (Fig. 4D). This result indicates that the electronic perturbation caused by a nitrogen dopant is localized near the dopant atom, which is confirmed in large-area dI/dV maps, and seen in the calculated charge distribution [see SOM II(e) and figure S5] and simulated STM image in Fig. 2B (*26*).

39. This material is based upon work supported as part of the Center for Re-Defining Photovoltaic Efficiency Through Molecule Scale Control, an Energy Frontier Research Center funded by the U.S. Department of Energy (DOE), Office of Science, Office of Basic Energy Sciences under Award Number DE-SC0001085. Support also provided by the Air Force Office of Scientific Research under grants FA9550-11-1-0010 (A.N.P); by the DOE under grant no. DE-FG02-88ER13937 (G.W.F) and grant no. DE-FG02-07ER15842 (T.H.) for research carried out in part at the Center for Functional Nanomaterials, Brookhaven National Laboratory, contract no. DE-AC02-98CH10886 (M.S.H.) and at the National Synchrotron Light Source, contract no. DE-AC02-98CH10886; by the Office of Naval Research under Graphene Multidisciplinary University Research Initiative (A.P. and P.K.); by Defense Advanced Research Projects Agency Carbon Electronics for RF Applications program (P.K.); by the National Science Foundation under grant no. CHE-0641523 (A.P.); by New York State Office of Science, Technology and Academic Research, and by Priority Research Centers Program (2011-0018395) through the National Research Foundation of Korea funded by the Ministry of Education, Science and Technology (K.S.K.). Equipment and material support was provided by the NSF under grant CHE-07-01483 (G.W.F.). Portions of this research were carried out at the Stanford Synchrotron Radiation Lightsource (SSRL), a Directorate of SLAC National Accelerator Laboratory and an Office of Science User Facility operated for the U.S. DOE Office of Science by Stanford University. We thank Cherno Jaye and Daniel Fischer for assistance in using National Synchrotron Light Source beamline U7A, Hirohito Ogasawara for assistance at SSRL beamline 13-2, and Chris Marianetti and Deborah Prezzi for useful discussions. The authors declare no competing financial interests. Requests for materials should be addressed to A.N.P.

**Figure Captions**

**Figure 1 Raman, XPS and STM of N-doped graphene (A)** Raman spectra taken at pristine and N-doped (four different doping levels) graphene on a SiO$_2$/Si substrate (20 um laser spot) showing systematic changes in the spectra as a function of doping. **(B)** NEXAFS (total electron yield) of pristine and N-doped graphene on copper foil at the N K-edge. N-doping results in a new peak in the spectrum at 400.7 eV due to graphitic N. XAS, x-ray absorption spectroscopy. (Inset) XPS data for pristine and NG10 graphene, showing a higher binding energy component (black arrow) for the doped sample. **(C)** Large area STM image of N-doped graphene on copper foil showing the presence of numerous point-like dopants(such as the dashed box) and occasional clusters of dopants (indicated by triangles) ($V_{bias}$=0.8V, $I_{set}$=0.8nA).

**Figure 2 STM imaging of nitrogen dopants (A)** STM image of the most common doping form observed on N-doped graphene on copper foil, corresponding to a single graphitic N dopant. (Inset) Line profile across the dopant shows atomic corrugation and apparent height of the dopant ($V_{bias}$=0.8V, $I_{set}$=0.8nA) **(B)** Simulated STM image of graphitic N dopant ($V_{bias}$=0.5 V), based on DFT calculations. Also superposed is a ball-and-stick model of the graphene lattice with a single N impurity. **(C)** STM image of N-doped graphene on copper foil showing 14 graphitic dopants and strong inter-valley scattering 'tails'. (Inset) FFT of topography shows atomic peaks (outer hexagon) and inter-valley scattering peaks(inner hexagon, indicated by red arrow)($V_{bias}$=0.8V, $I_{set}$=0.8nA)**(D)** Spatial distribution of N-N distances from eight samples on copper foils with different N concentrations. The distributions are all fit well by a quadratic power law (expected error bands in gray) overall length scales indicating that N dopants incorporate randomly into the graphene lattice.**(E)**dI/dV curves taken on a N atom (bottom) and on the bright topographic features near the nitrogen atom on N-doped graphene on copper, offset vertically for clarity. The top curve is the dI/dV spectrum taken at ~2nm away from the dopant. (Inset) Positions where the spectra were taken.($V_{bias}$=0.8V, $I_{set}$=1.0nA).

**Figure 3 Spectroscopy and doping in N-doped graphene (A)** Histogram of the spatial variation of the Dirac point for N-doped graphene on copper foil over an area of 40nm by 40nm ($V_{bias}$=0.6V, $I_{set}$=1.0nA). (Inset) Spatially averaged dI/dV spectrums (black line) and variation (gray band) over the area. **(B)**Spatially-averaged graphene charge carrier concentration as a function of average nitrogen concentration level for five different samples measured on N-doped graphene on copper foil. (Inset) Free charge carriers per nitrogen atom for each of these samples. **(C)** Calculated projected density of states near the Fermi energy for a 1% doping level. **(D)** Averaged dI/dV spectrum for a N-doped graphene transferred to a SiO$_2$/Si substrate (black line) and spatial variation (gray band). The Dirac point is at -330meV (red arrow) ($V_{bias}$=1.0V, $I_{set}$=0.1nA). (Inset) Topography shown in the derivative mode in order to enhance atomic contrast.

**Figure 4 Spectroscopic mapping around a single N dopant (A)** dI/dV spetra taken on the N atom(solid line) and far away from the N atom (dashed line) for N-doped graphene on copper foil ($V_{bias}$=0.8V, $I_{set}$=1.0nA)**(B)** STM spectroscopic maps taken in the vicinity of a single N dopant for N-doped graphene on copper foil (dopant position indicated by a dot) at different energies with N positions marked by red diamonds ($V_{bias}$=0.8V, $I_{set}$=1.0nA)**(C)** Radially-averaged differential conductance as a function of the distance from the N atom at different energies, normalized to unity at distances far away from the N atom. The fits are to an exponential decay function. **(D)** Extracted decay lengths for different energies (square) and the ratio of the differential conductance on the N site to the background (triangles) for different energies. The decay lengths are ~7Å for all energies.

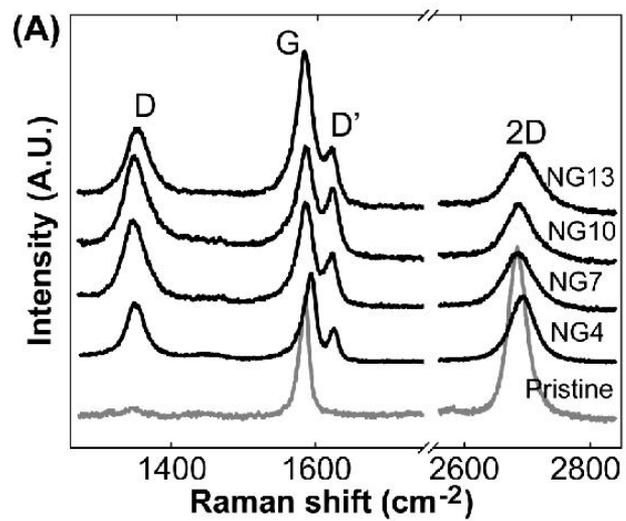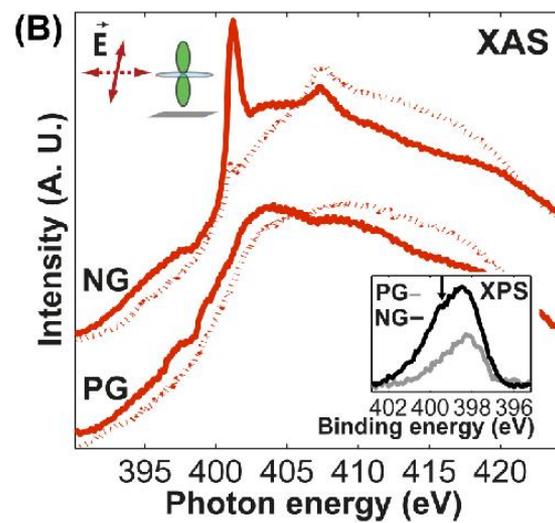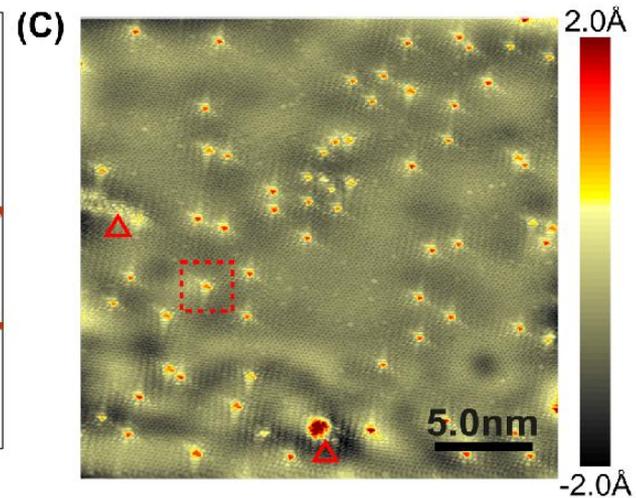

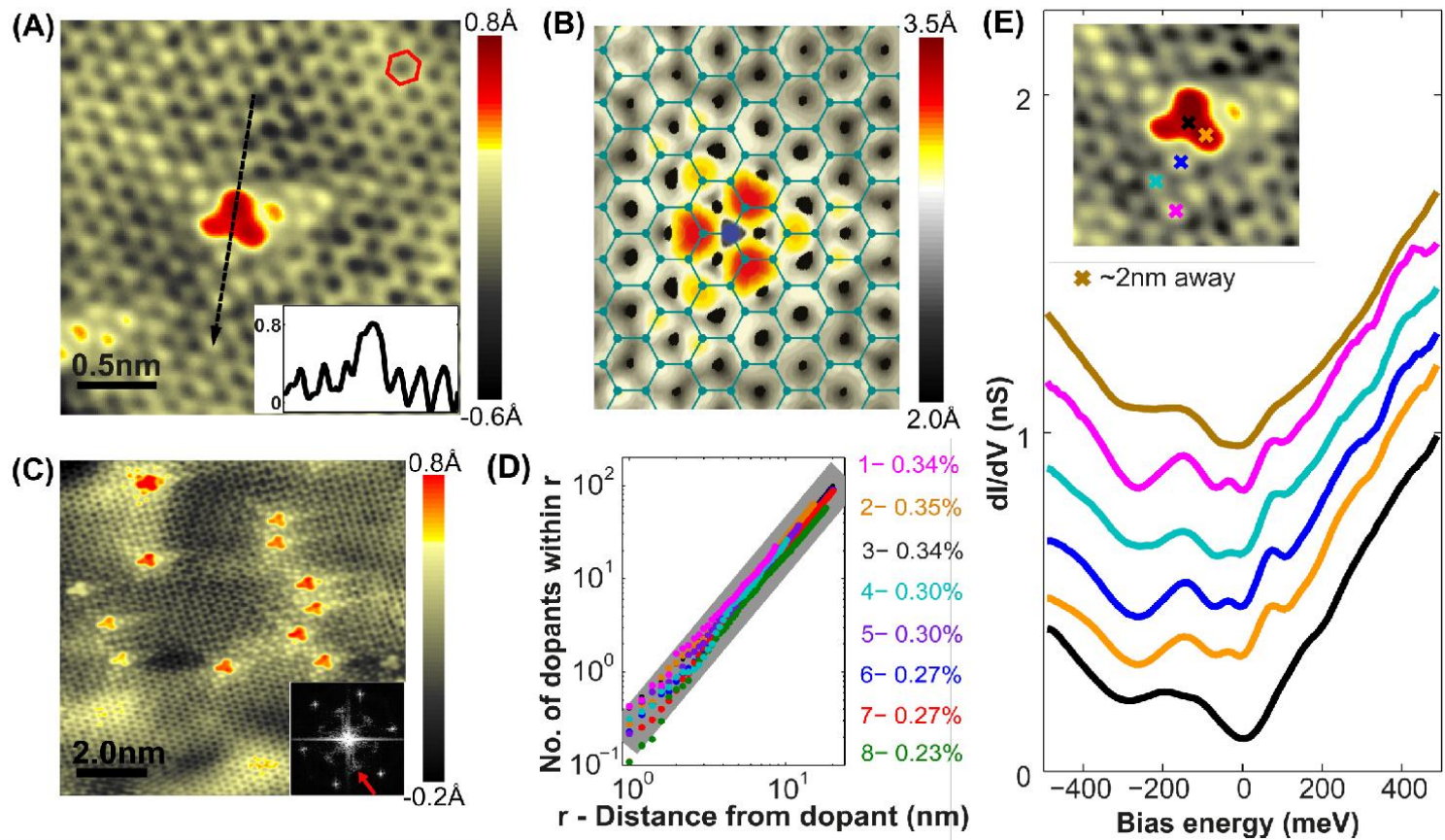

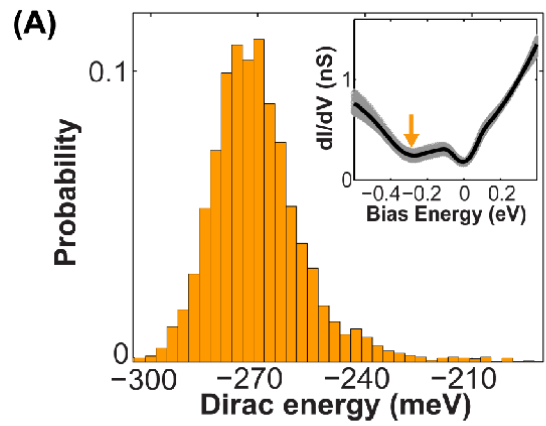
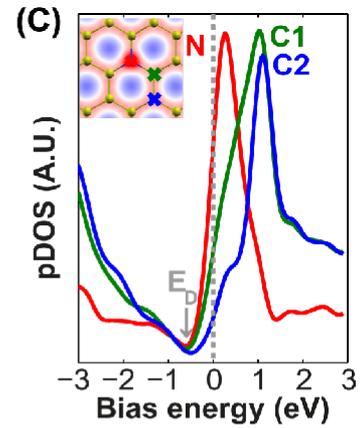
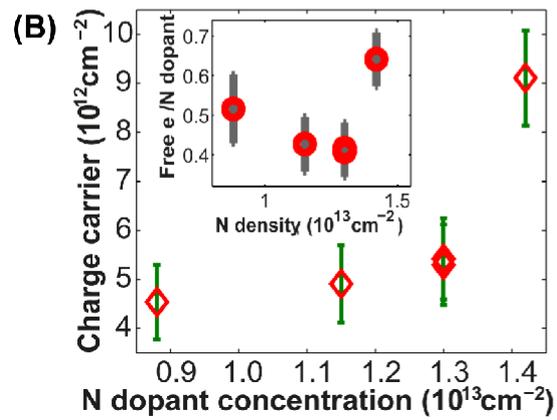
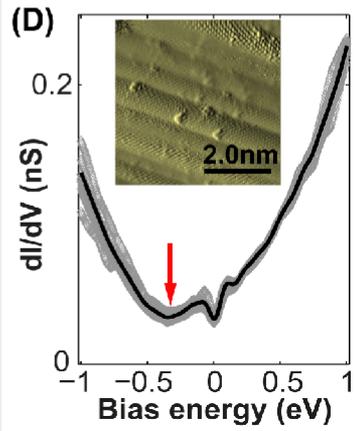

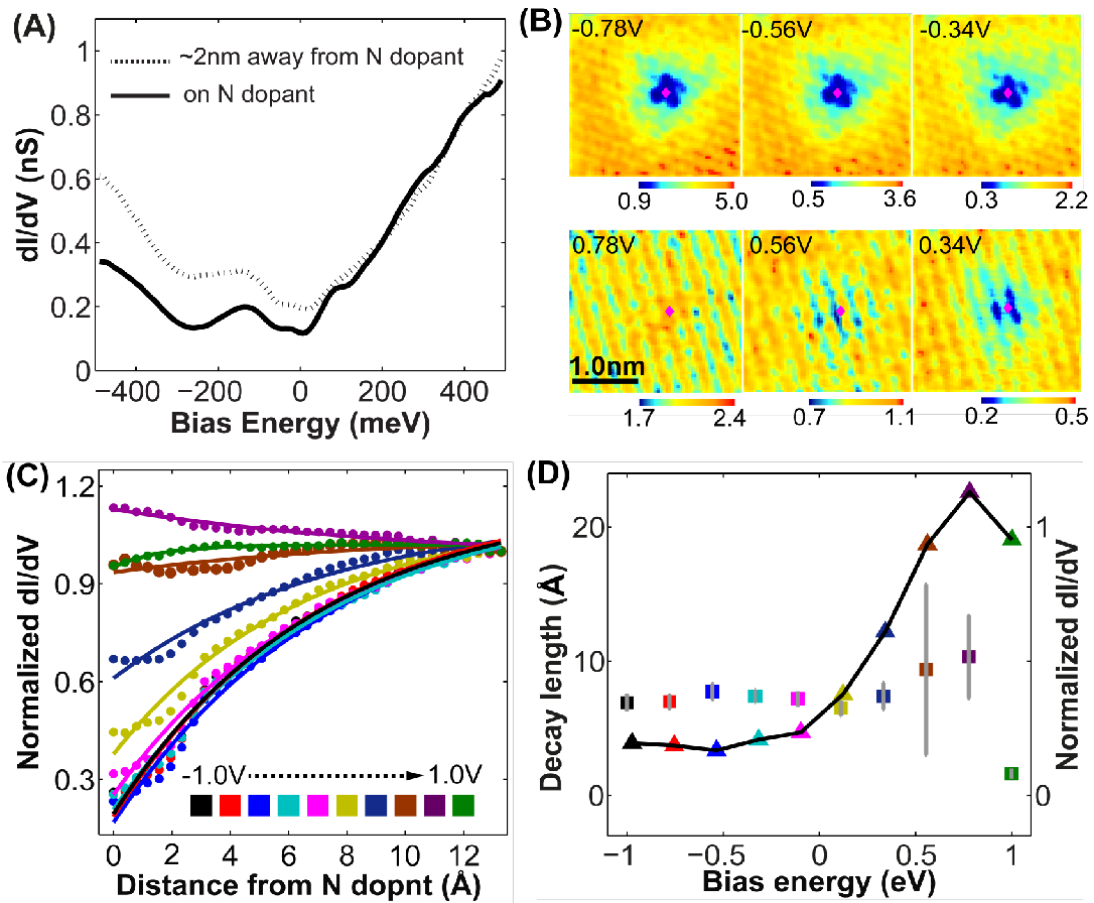

# Visualizing Individual Nitrogen Dopants in Monolayer Graphene

L. Zhao[1], R. He[1], K. T. Rim[4], K. S. Kim[1], H. Zhou[1], C. Gutiérrez[1], S. Chockalingam[1], C. Arguello[1], L. Pálová[4], T. Schiros[5], D. Nordlund[6], M.S. Hybertsen[7], D.R. Reichman[4], T. F. Heinz[1,2], P. Kim[1], A. Pinczuk[1,3], G. W. Flynn[4], A. N. Pasupathy[1,*]

[1]Department of Physics, Columbia University, New York NY 10027
[2]Department of Electrical Engineering, Columbia University, New York NY 10027
[3]Department of Applied Physics and Applied Mathematics, Columbia University, New York NY 10027
[4]Department of Chemistry, Columbia University, New York NY 10027
[5]Energy Frontier Research Center, Columbia University, New York NY 10027
[6]Stanford Synchrotron Radiation Laboratory, Menlo Park, CA 94025
[7]Center for Functional Nanomaterials, Brookhaven National Laboratory, Upton, New York, 11973-5000

*Contact: apn2108@columbia.edu

**I. Materials and Methods:**

**(a) Preparation of graphene films by chemical vapor deposition**

The foil substrate was precleaned with a flow of 10 sccm of $H_2$ at a pressure of 0.055 torr and a temperature of 1000 °C for 10 min. Doped graphene films were then grown using a mixture of $CH_4$, $H_2$ and $NH_3$ at a total pressure of 1.9 torr and a temperature of 1000 °C for 18 min (*12*). While all the samples were grown with $CH_4$ and $H_2$ flow rates of 170 sccm and 10 sccm respectively, different doping concentrations were achieved using $NH_3$ partial pressures of 0 torr (pristine graphene, PG), 0.04 torr (NG4), 0.07 torr (NG7), 0.10 torr (NG10) and 0.13 torr (NG13).

**(b) Film Transfer to $SiO_2$ substrates**

Raman, X-Ray and STM measurements were performed on graphene films both on copper foil substrates as well as $SiO_2$ on Silicon substrates. To transfer the as-grown graphene films to $SiO_2$/Si substrates, the graphene films on copper were first coated with a 200 nm layer of PMMA. The underlying copper was then etched in $FeCl_3$ (20 wt%) solution till it was completely dissolved. The graphene/PMMA layer floats on the solution and can be picked up using a $SiO_2$/Si substrate. Several rinses in deionized water are performed before allowing the substrate to dry in air.

**(c) Sample preparation for NEXAFS experiments**

Graphene-coated substrates were placed on a sample holder and contacted using copper or carbon tape. All measurements were performed at room temperature in UHV conditions.

**(d) Sample preparation for STM experiments**

Graphene-coated substrates were placed on a sample holder and tantalum strips were used to make electrical contact to the large-area graphene films, both for the copper substrate as well as the $SiO_2$/Si substrate. The samples were degassed in the UHV measurement chamber at 350$^o$ C for 5 hours before measurements were performed.

**(e) X-ray spectroscopy experimental details**

NEXAFS and XPS measurements were performed at Stanford Synchrotron Radiation Lightsource (SSRL) beamlines 10-1 and 13-2; NEXAFS was also measured at the NIST beamline U7A of the National Synchrotron Light Source (NSLS). The reference absorption intensity ($I_0$) of the incoming x-ray beam, measured on a gold coated mesh positioned just after the refocusing optics, was measured simultaneously and used to normalize the spectra to avoid any artifacts due to beam instability. Polarization dependent NEXAFS data was obtained by either changing the angle between the incoming x-ray beam from near parallel (10°(SSRL) and 20° (NSLS)) and near normal (90°(SSRL 10-1) and 70° (NSLS)) incidence or by changing the polarization of the electric field vector between in- and out-of-plane geometries at EPU beamline 13-2 (SSRL). Nitrogen K-edge XAS shown here was measured in total electron yield (TEY) mode using 30x50 um slits. A linear background was subtracted from a region before the absorption edge (385-392 eV). Spectra were normalized by area with respect to nitrogen concentration using a two-point normalization: area normalization between 392-425 eV and a continuum normalization in the region 425-430 eV (atomic normalization).

SSRL beamlines 13-2 and 10-1 have spherical grating monochromators while NSLS U7A utilizes a toroidal spherical grating monochromator; the focused beams at 13-2, 10-1, and U7A have spot sizes of 0.01 x 0.075 mm$^2$, < 1mm$^2$, and ~1.5 mm$^2$, respectively. All three end stations are designed for surface and solid state experiments with ultra-high vacuum compatible samples. XPS spectra were measured at beamline 13-2 with a hemispherical electron spectrometer (SES-R3000, VG-Scienta) using a total energy resolution better than 100 meV. The XPS binding energy scale for spectra measured at photon energy 600 eV was calibrated using the shift between the monochromator at 310 eV and actual energy (determined by higher order photoemission lines of the graphene C1s peak).

**II. SOM text:**

**(a) Raman measurements on copper substrates versus $SiO_2$ substrates**

Raman measurements have been performed on the doped graphene samples both on copper substrates and $SiO_2$ substrates. Shown below in figure S1 are the spectra obtained for both substrates for various doping levels. The main features of the spectra including the shift of the G band and the intensity changes in the D and 2D bands are observed on both substrates. Spectra obtained on the copper substrate show an additional background due to substrate luminescence.

**(b) Other doping forms seen in STM experiments**

While most of the dopants that are observed are in the graphitic substitutional form, we occasionally see clusters of two or more dopants. Shown in figure S2 are four different such doping forms. Since STM measures only the local density of states and lacks chemical specificity, we cannot conclusively identify the nature of these dopant configurations without detailed theoretical modeling (currently in progress).

**(c) Incorporation of dopants on the two sublattices of graphene**

A given graphitic dopant can exist on one of the two sublattices of graphene. It is interesting to ask whether there is a local preference of dopants to incorporate on the same sublattice. Indeed, some visual evidence is seen for this in large-area STM topographs such as the ones shown in figure S3. We see that there are local patches where a majority of dopants tend to incorporate on the same sublattice. Our current data is limited by the size of the flat and clean areas seen in the topography of the copper foil. Experiments on ultraflat copper single crystals (currently in progress) can allow us to scan large enough areas to obtain statistically large data sets to investigate this phenomenon.

**(d) Scanning Raman Measurements**

While STM is a powerful technique with which to extract the information about electronic structure and doping on nanoscale, it is also important to characterize the electronic homogeneity of the films on the micrometer scale. To address this issue, we performed scanning Raman spectroscopy measurements to probe the homogeneity of nitrogen doping by the CVD process. Shown in figure S4A are typical Raman spectra taken at different locations on the same sample as used in the STM studies. All the spectra show the G and 2D bands for pristine graphene , as well as the D and D' bands that are seen in the presence of defects. However, we see that different locations of the sample exhibit very different peak magnitudes and positions. In particular, some spectra display characteristics of graphene with a low carrier/defect density (bottom curve - high 2D/G peak ratio, small D/G and D'/G peak ratio), while other spectra show the presence of more dopants (top curve). To better understand the spatial inhomogeneity, we performed these spectroscopic measurements at every location of the PG, NG10 and NG13 samples over a 80×80 $\mu m^2$ area. Shown in figure S4B are maps of the ratio of the 2D/G peak heights, extracted from the spectral maps. In previous experiments on pristine graphene, this ratio was shown to be sensitive to the carrier and defect concentration in the film. We see firstly that the PG sample shows a much higher 2D/G ratio on average, indicating fewer defects or dopants in the film. The doped samples clearly show the presence of micrometer-sized patches with higher nitrogen concentration (purple). The boundaries of these patches (yellow) correspond to a large 2D/G ratio (~2), indicating a lower nitrogen concentration than the interior of the patches. We see that the size of the patches is doping dependent, indicating that the patches are truly related to the local dopant concentration. These conclusions are supported by the analysis of the G-mode peak energy - in previous measurements of N-doped graphene, the position of the G band was correlated with the amount of nitrogen present in the sample (*14*). Shown in figure S4C are maps of the G peak position in the same area of the samples. Histograms of these peak positions for each sample are shown in figure S4D. We see a clear shift in the position of the G peak across samples indicating that the carrier concentration in the graphene film depends on the nitrogen concentration. The size of the shift between the PG and NG10 sample (2.0 $cm^{-1}$) is consistent with our STM measurements of Dirac point position in the doped film.

**(e) DFT calculations of the charge density around dopants**

First-principles calculations are performed using density functional theory within the local density approximation (LDA, Perdew-Zunger fit of the Ceperley-Alder electron gas results) for exchange and correlation as implemented in the Quantum Espresso package (*28*). Norm-conserving pseudopotentials provided with the package (Von Barth-Car direct-fit form with LDA) are used with 5 valence electrons for nitrogen and 4 valence electrons for carbon. The cutoff energy for the plane-wave basis set is 95 Ry. The calculated lattice constant for pristine graphene is 2.44 Å. Graphene layers are separated by 14.64 Å vacuum to form a supercell. We introduce nitrogen into the graphene lattice and we perform structural relaxation until the forces are less than $10^{-3}$ Ry/a.u. We explicitly checked to see that the N remains coplanar with the surrounding graphene. We use several hexagonal supercells in order to treat different nitrogen doping levels: 3x3, 4x4, 5x5, 6x6, 7x7, 8x8 and 9x9 supercells are used, containing 18, 32, 50, 72, 98, 128 and 162 atoms, respectively, which correspond to 5.6 - 0.6% nitrogen doping. Uniform 9x9, 7x7, 5x5, 4x4 and 3x3 Monkhorst-type k-point grids with zero offsets are generated for the 3x3, 4x4, 5x5, 6x6 (7x7), and 8x8 (9x9) supercells respectively for self-consistency and structural relaxation. The nitrogen-carbon nearest neighbor distance is found to be 1.40 Å for the graphitic form of doping, with less than 0.002 Å change with cell size (N concentration) indicating very localized structural changes due to the substituent. This is confirmed by the localized perturbation to the electronic charge density, illustrated in figure S5. For reference, the charge density is also shown for two other, relaxed N structures.

Our STM image simulations have been carried out using the Tersoff-Hamann approach (*27*). We apply a bias voltage of +0.5 eV, hence we probe the nitrogen-doped graphene empty states. The projected density of states calculations have been performed using 18x18 (for the 3x3), 12x12 (for the 4x4), 9x9 (for the 5x5 and 6x6) and 6x6 (for the 7x7, 8x8 and 9x9 graphene supercells) Monkhorst-type k-point grids with zero offsets. The STM image simulation in Fig. 2B and the pDOS in Fig. 3C were calculated for the 7x7 supercell. Following the experimental protocol, the value of the Dirac point has been evaluated from the minimum in the total density of states. Using the same formula and parameters described in the text, we deduce that the free carrier concentration varies between 11.4 (for the 3x3) to $1.7 \times 10^{13}$ cm$^{-2}$ (for the 9x9 supercell). As shown in figure S6 below, the ratio of the deduced free carrier concentration to the N concentration exhibits a clear, period 3 oscillation that is commonly observed for many phenomena in graphene where coherent electronic interference effects play a role. In addition, the ratio is somewhat lower for the small cells where overlap of the perturbation caused by N atoms in neighboring cells plays a role. In the samples measured in experiment, the N atoms are randomly placed, so the role of electronic interference should be minimal. We used an average of the largest three cells to have an estimate: 61%.

**(f) STM measurement of pristine graphene (PG) films on copper foil**

In order to see the charge transfer between the copper foil and graphene, we performed STM measurements of pristine graphene films on copper foils. Shown in figure S7 is the spatially averaged spectrum of pristine on copper foil. The topography of the sample shown in the inset shows the absence of dopant features and only shows a weak Moire pattern (*40*) We see that the spectrum of pristine graphene shows only one minimum near the Fermi energy, which shows a doping level < $10^{12}$ cm$^{-2}$.

SOM-5

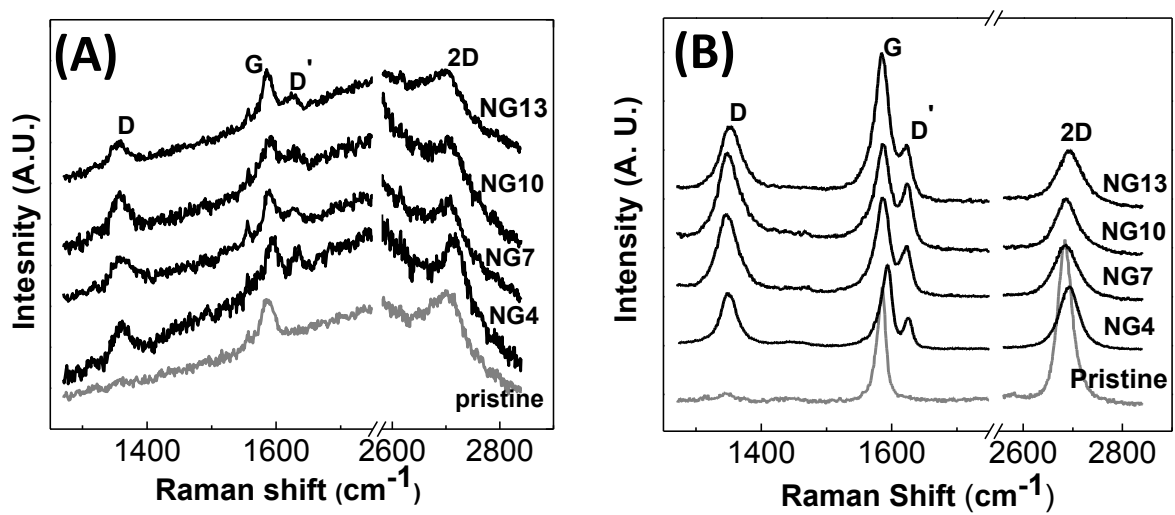

figure S1: (A) Raman spectra taken on as-grown pristine and N-doped graphene on copper foil (4 different doping levels), showing a background due to copper luminescence. (B)Raman spectra taken for the same samples in (A) after transferring to SiO$_2$/Si substrates.

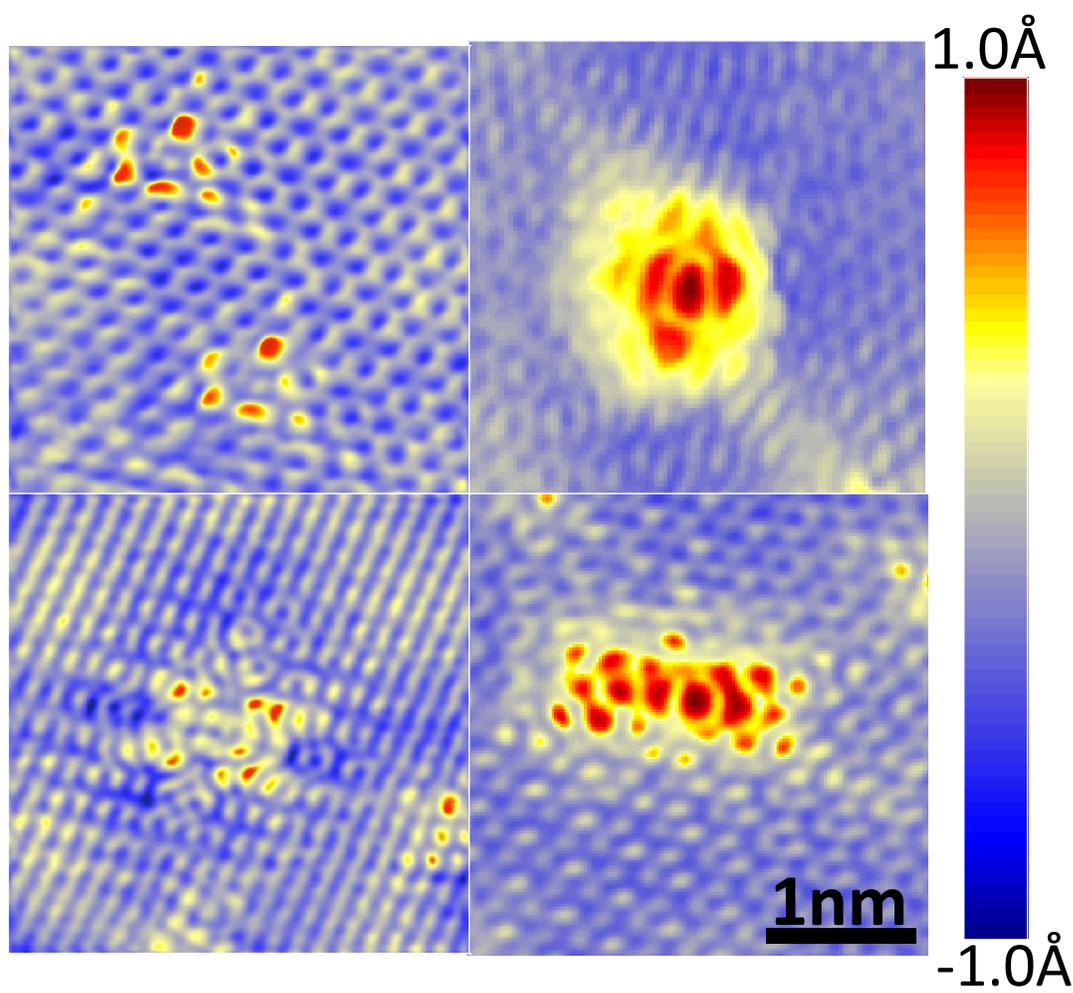

figure S2: STM images of four different dopant structures found on NG10 graphene films

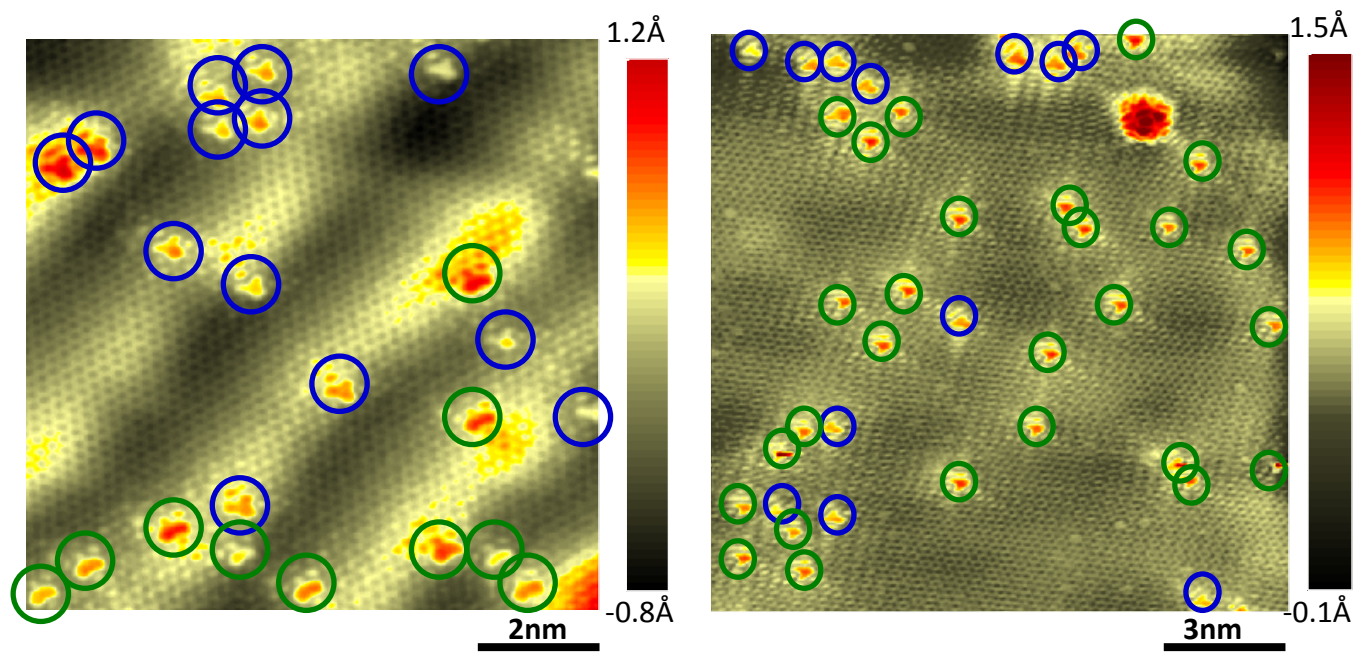

figure. S3: Single substitutional N dopants can incorporate in either sublattice of graphene (indicated by blue and green circles). We see some visual evidence for dopants to choose the same sublattice at local (few nm) scales.

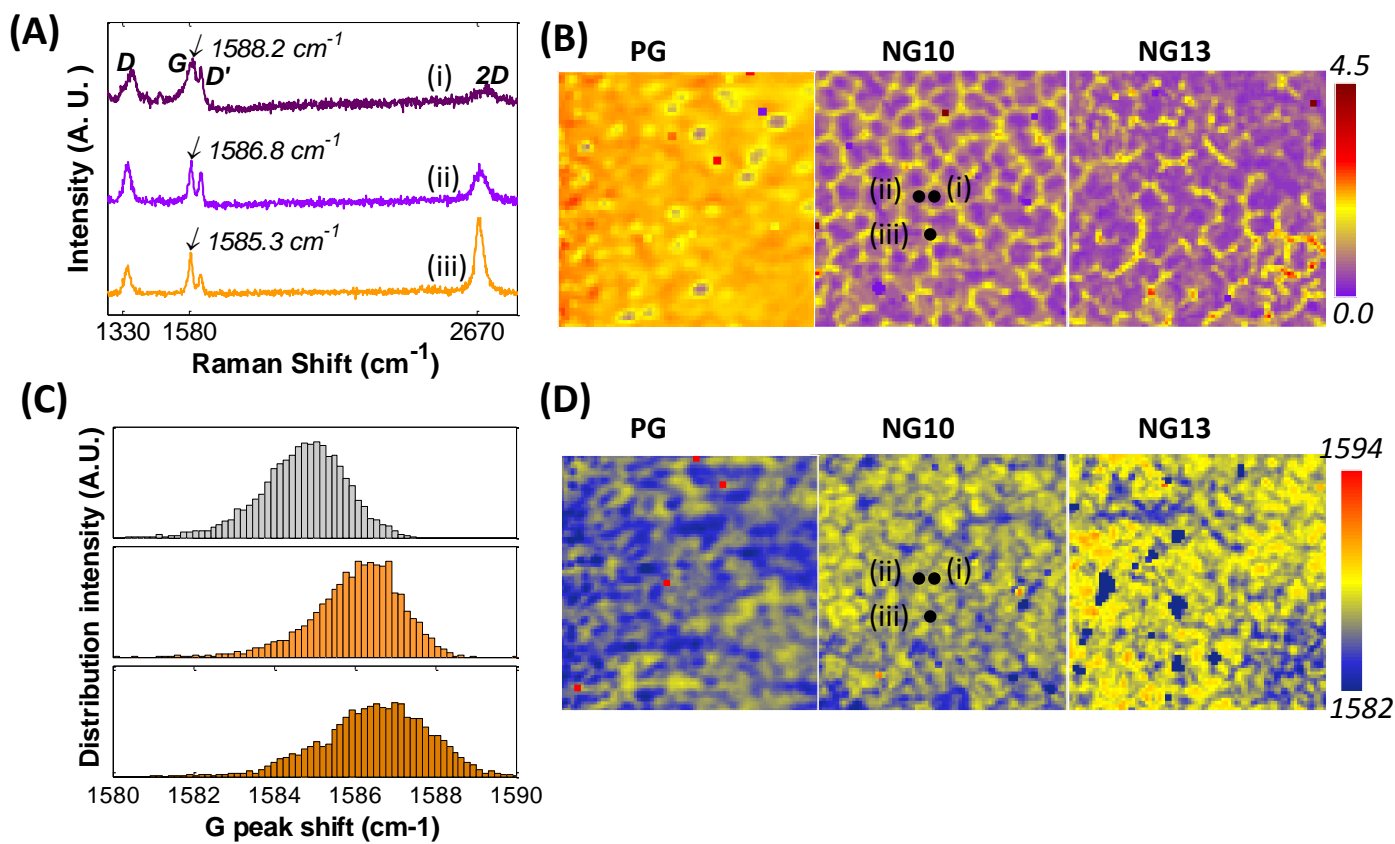

figure S4: (A) Raman spectra taken at different positions over an NG10 sample surface after transferring to SiO$_2$/Si. The positions where the spectra were taken are indicated in the NG10 sample in (B) and (D). (B) Maps of the ratio of the 2D/G peak intensities for PG, NG10 and NG13 samples. (C) Maps of the G peak frequency for PG, NG10 and NG13 samples. (D) G peak frequency histogram of PG, NG10, NG13 from the maps in (C).

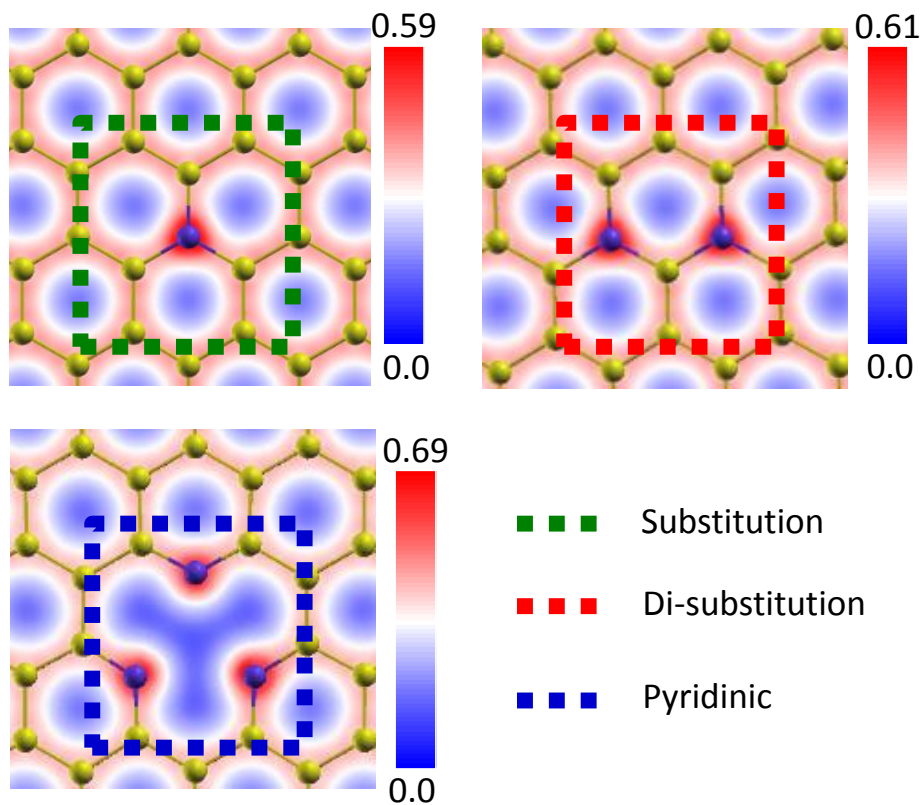

figure S5: DFT calculations of the charge density maps for three doping forms. Models of doping forms are laid on top of charge density maps, with blue spots for nitrogen dopants and yellow spots for carbon atoms.

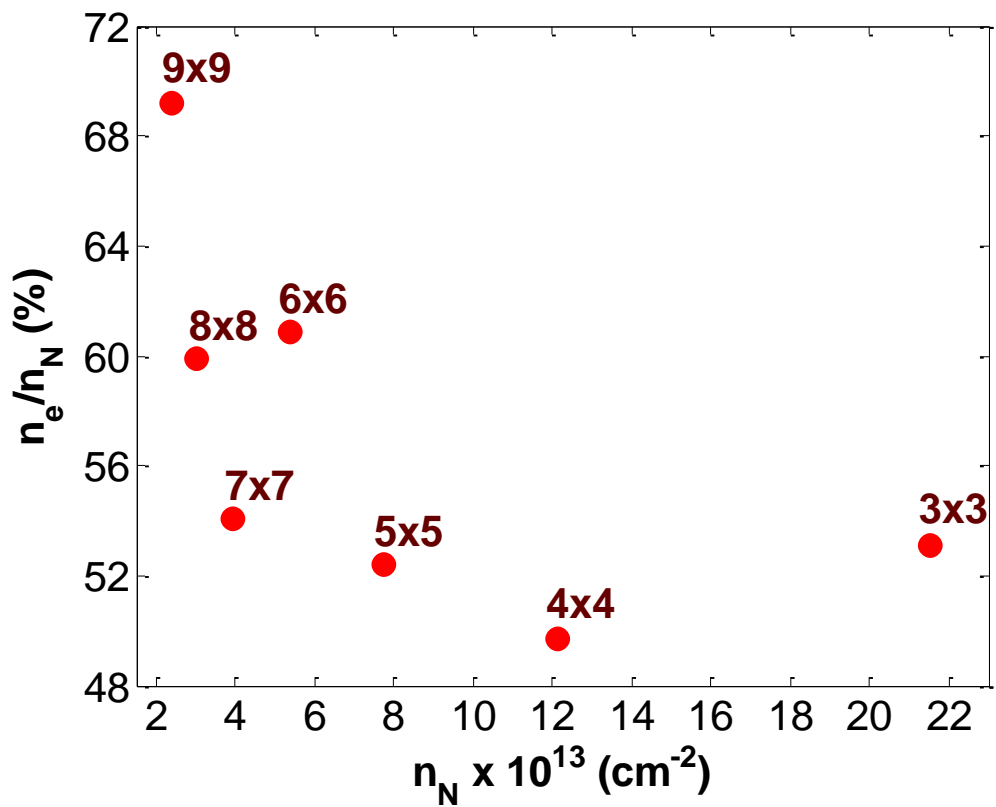

figure S6: DFT calculation of the ratio of the free charge carrier concentration to the N concentration. Each point is indicated with the unit cell used for DFT calculation.

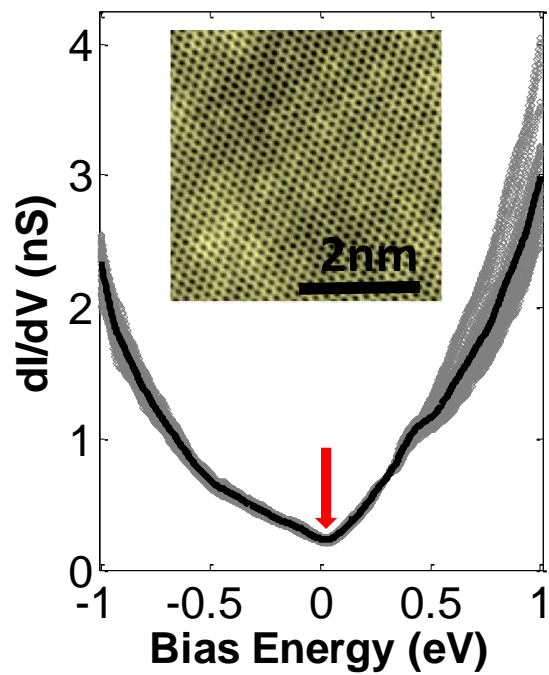

figure S7: Averaged dI/dV spectrum (black line) and spatial variation (gray band) for a pristine graphene sample on copper foil. The Dirac point is near zero bias. The topography of the area is shown in the inset and shows the absence of dopant structures ($V_{bias}$=0.6V, $I_{set}$=0.7nA).